%% file: sample-sigconf.tex
\documentclass[sigconf]{acmart}[review]
\usepackage{soul}
\usepackage[normalem]{ulem}
\usepackage{booktabs} 
\usepackage{balance}
\usepackage{epsfig}
\usepackage{subfigure}
\usepackage{calc}
\usepackage{natbib}
\usepackage{multirow}
\usepackage{xcolor}
\usepackage{amsmath}
\usepackage{listings}
\usepackage{enumitem}
\usepackage{graphicx}
\usepackage{multirow}
\usepackage{longtable}
\usepackage{tcolorbox}
\usepackage{supertabular}
\usepackage{tabularx}

\usepackage{fancyhdr}

\setcopyright{none}
\fancypagestyle{FirstPage}{
\cfoot{\textit{\small\copyright IEEE.  Personal use of this material is permitted.  Permission from IEEE must be obtained for all other uses, in any current or future media, including reprinting/republishing this material for advertising or promotional purposes, creating new collective works, for resale or redistribution to servers or lists, or reuse of any copyrighted component of this work in other works.}} 
}

\begin{document}
\copyrightyear{}
\acmYear{}
\acmConference[\textbf{\textit{NOTE: }} This manuscript is a pre-publication version of the paper that was published at IEEE SEAA 2024, doi: 10.1109/SEAA64295.2024.00050~~~]{}{}{}

\title{Measuring Software Development Waste \\in Open-Source Software Projects}

    
\author{Dhiraj SM Varanasi}
\affiliation{
  \institution{SERC, IIIT Hyderabad}
}
\email{dhiraj.shanmukha@research.iiit.ac.in}

\author{Divij D}
\affiliation{
  \institution{SERC, IIIT Hyderabad, India}
}
\email{divij.d@students.iiit.ac.in}

\author{Sai Anirudh Karre}
\affiliation{
  \institution{SERC, IIIT Hyderabad, India}
}
\email{saianirudh.karri@research.iiit.ac.in}

\author{Y Raghu Reddy}
\affiliation{
  {\institution{SERC, IIIT Hyderabad, India}}
}
\email{raghu.reddy@research.iiit.ac.in}

\begin{abstract}
Software Development Waste (SDW) is defined as any resource-consuming activity that does not add value to the client or the organization developing the software. SDW impacts the overall efficiency and productivity of a software project as the scale and size of the project grows. Although engineering leaders usually put in effort to minimize waste, the lack of definitive measures to track and manage SDW is a cause of concern. To address this gap, we propose five measures, namely \textit{Stale Forks}, \textit{Project Diversification Index}, \textit{PR Rejection Rate}, \textit{Backlog Inversion Index}, and \textit{Feature Fulfillment Rate} to potentially identify \textit{unused artifacts}, \textit{building the wrong feature/product}, \textit{mismanagement of backlog} types of SDW. We apply these measures on ten open-source projects and share our observations to apply them in practice for managing SDW. 
\end{abstract}


\maketitle
\renewcommand{\shortauthors}{Varanasi et al.}
\input{samplebody-conf}
\balance
\bibliographystyle{ACM-Reference-Format}
\bibliography{Main-Bibliography}
\end{document}

%% file: samplebody-conf.tex
\section{Introduction}
\thispagestyle{FirstPage}
\label{section:introduction}
Software development involves activities that deal with creating, designing, deploying, and maintaining software. In the context of software development, waste refers to any activity, process, or artifact that consumes resources but fails to add value to the end product or service. Software development waste (SDW) manifests in various forms, such as building the wrong feature or product, mismanaging the backlog, producing unused artifacts, engaging in ineffective communication, and other process inefficiencies. The cause may be due to people, process or the domain itself. Moreover, if not addressed, waste can evolve and impact other aspects of the software and lead to productivity issues \cite{varanasi2024software}\cite{sedano2017software}. Identifying and eliminating SDW is paramount for enhancing productivity, reducing costs, and ensuring delivery timelines. For example, consider a large banking project with multiple teams working on different modules. Similar to any large software, integrating the modules requires a thorough quality check, any amount of rework or building wrong features impacts the project's overall delivery. Detecting and tracking SDW is critical for software product delivery as it avoids potential productivity issues and slippage delivery timelines. 

Organizations are increasingly adopting open-source software (OSS) instead of proprietary software as its considered to be of high quality  \cite{taibi2015empirical} and to reap the benefits of innovation and resource management \cite{hauge2010adoption}. Given the increase in usage of OSS\cite{linaaker2023open}, it is prudent to understand the causes of SDW in OSS, and the metrics to be used for their measurement. As part of this paper, we examine the state-of-the-art practices for measuring software development waste in general and more specifically OSS. We  identify the gaps in the existing measures and propose new measures to calculate SDW. The paper focuses on the following research questions:

\begin{tcolorbox}
\begin{itemize}
    \item [\textbf{RQ1: }] What are the known types of software development waste? 
    \item [\textbf{RQ2: }] How are these software development wastes measured? 
    \item [\textbf{RQ3: }] How much software development waste(s) occur in Open-Source software projects? 
\end{itemize}
\end{tcolorbox}

RQ1 examines the types of software development waste(s). RQ2 is an extension to RQ1 and explores how the identified SDW are measured. RQ3 analyzes the occurrence of SDW in open-source software projects. The combination of questions helps us utilize the measures identified in RQ2 and propose any additional measures that may be needed to assess SDW in open-source software.

While SDW remains an area of interest for industry \cite{varanasi2024software}, there is negligible work on quantifying SDW. Kai Petersen \cite{petersen2012palette} and Alahyari et al. \cite{alahyari2019exploratory}, Sallin et al. \cite{sallin2023waste} suggest using lean  indicators like lead time and defect counts as indirect means to assess SDW. This gap motivated us to formulate new measure for assessment of SDW.

In this paper, the following proposed measures and the corresponding SDW they are associated are detailed:  (1) \textit{`stale forks'} for \textit{`unused artifacts'}, (2) \textit{`project diversification index'} for \textit{`building the wrong feature/product'}, (3) \textit{`pr rejection rate'} for \textit{`unused artifacts'} and \textit{`mismanagement of backlog'}, and (4) \textit{`backlog inversion index'} and (5) \textit{`feature fulfillment rate'} for \textit{`mismanagement of backlog'}. These measures can potentially assist teams to make decisions and improve the quality of the final product. Software development teams can use the proposed measures to assess SDW and take corrective actions to minimize SDW.  

\section{Related Work}
\label{section:related_work}






Sedano et al. \cite{sedano2017software} and Alahyari et al. \cite{alahyari2019exploratory} provided a comprehensive list of SDW through elaborate participant-observation studies, identifying common waste types such as rework, waiting, and ineffective communication and more and mentioned the lack of underscoring unified approach to defining and measuring waste \cite{alahyari2019exploratory}. While studies have not explicitly mentioned any SDW measures, Kai Petersen presents lean indicators such as \textit{maintenance request inflow}, \textit{lead time}, and \textit{workload} to identify waste and uses the \textit{cumulative flow diagram to visualize handovers} \cite{petersen2012palette}. \textit{Number of Accumulated Items} and the \textit{Work in Progress} are the unfinished artifacts that are not providing value to the customer at the time \cite{bufon2019method}. Alahyari et al. mentions \textit{defect}, \textit{lead-time}, \textit{timely delivery}, \textit{team velocity}, and \textit{trouble reports} as measurements for waste \cite{alahyari2019exploratory}. 

With a much more explicit focus on the quantification of specific types of waste, Sallin et al. \cite{sallin2023waste} expanded on the idea of self-reporting waste in terms of time spent and delays for mismanaging the backlog, rework, unnecessarily complex solutions, extraneous cognitive load, knowledge loss, ineffective communication, management \& organizational aspect, manual work, other duties. Building the wrong feature/product was measured through customer confidence on a Likert scale, psychological distress with stress on a numerical scale, and waiting or multitasking through delays and context switching. Despite these efforts, quantifying and measuring SDW remains a challenge \cite{alahyari2019exploratory}\cite{varanasi2024software}.  

\section{Proposed Measures}
\label{section:measures}
Studies suggest that while various software development frameworks recognize waste, quantifying it remains an open problem \cite{alahyari2019exploratory}\cite{varanasi2024software}. Organizations currently use productivity indicators as a proxy for measuring SDW and are willing to adopt measures if they are easy to incorporate \cite{varanasi2024software}. We devised measures for \textit{unused artifacts}, \textit{mismanagement of backlog}, and \textit{building the wrong feature/product} types of SDWs, based on features available from open-source projects and discussions with SMEs. While there are other recognized SDWs such as \textit{psychological distress}, \textit{extraneous cognitive load}, and more, addressing all categories is beyond our scope. By focusing on these three, we provide a detailed, actionable analysis within our approach's constraints. This section discusses each measure and its expected outcome. Following are our proposed measures - \textit{Stale Forks, Project Diversification Index, PR Rejection Rate, Backlog Inversion Index, and Feature fulfilment Rate}
\vspace{-2pt}
\subsection{Stale Forks (SFs)}
\label{subsection:measure_stale_forks}


Forking is to create a copy of a repository to fix bugs, add features, create entirely new directions for the project, and more \cite{gousios2014exploratory}\cite{rastogi2016forking}. This practice is fundamental in open-source software development developers' exploration of the original vision without permission from the original repository maintainers. 

\begin{tcolorbox}
\textbf{Definition:} Copies of a software repository that do not add value to the project.\\ 
\noindent
\textbf{SDW Type:} Unused Artifacts \cite{alahyari2019exploratory}\cite{varanasi2024software} 
\end{tcolorbox}


\textit{`Stale Forks'} helps identify the \textit{`unused artifacts'} type of SDW by classifying the forks into subcategories - \textbf{`active'}, \textbf{`backup'}, \textbf{`potentially stale'} and \textbf{`stale.'} This classification can help the human in the loop further drill down on a \textit{`potentially stale'} forks to identify if they are \textit{`active'} or \textit{`stale'} and identify the \textbf{`unused artifacts'}. 

\begin{figure}[htbp]
\centering
\includegraphics[width=0.85\linewidth]{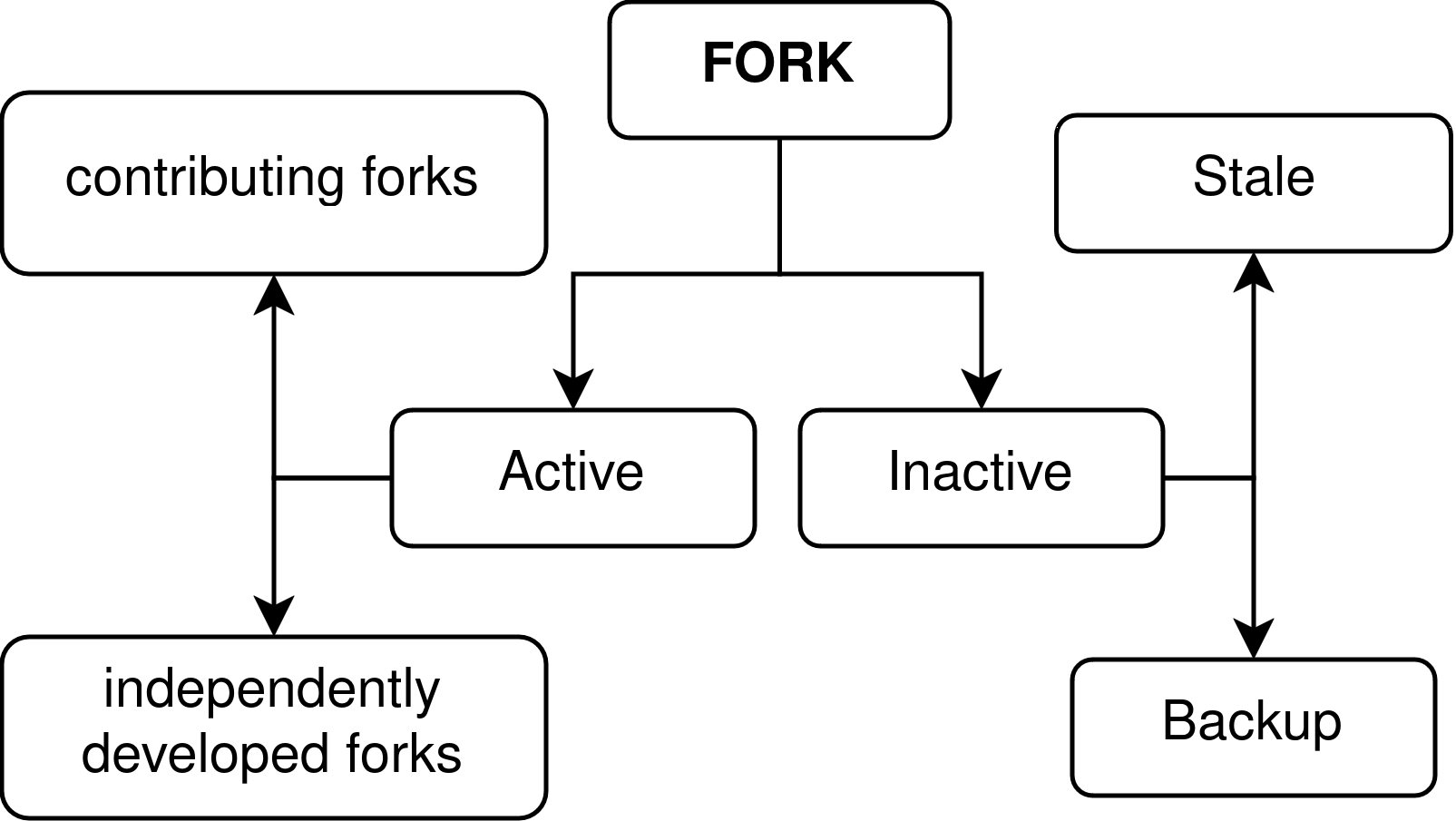} 
\caption{Types of Forks}
\label{fig:fork_types}
\end{figure}

\noindent
\textit{\textbf{Identifying Backup Forks:}} The scenario with forks having the \textbf{``pushed\_at"} timestamp preceding the \textbf{``created\_at"} timestamp indicates 
that no changes have been made since its inception. This pattern is a strong indicator that the created fork preserves the current state of the repository rather than for ongoing development or contribution back to the original project. Such forks, while not contributing to the active evolution of the software, play a crucial role in safeguarding the code base against potential loss or changes, thereby acting as a snapshot or backup of the project at a specific moment in its history. 
\begin{equation}
\textbf{\text{Backup Forks}} = \text{Fork}(\text{pushed\_at}) < \text{Fork}(\text{created\_at})
\label{eq:backup_forks}
\end{equation}
\noindent
\textit{\textbf{Identifying Active Forks:} } An \textbf{``active fork''} can be defined as one that has seen recent contributions in terms of commits, merges, or other forms of updates. For our study, we adopted GitHub's standard of sufficiency in data retention for 90 days sufficiency in data retention and classified the forks as active. The 90 days reflect GitHub's operational strategy of preserving detailed event logs for a limited duration. This time frame is optimal for maintaining efficient data management while capturing essential insights into the activities within the repository. Hence, we believe it is an appropriate time frame to classify forks as active. We classify the repositories with \textit{``pushed\_at''} not more than 90 days behind the \textit{``pushed\_at''} of the parent repository as active forks. 
\begin{equation}
\begin{split}
\textbf{\text{Active Forks}} &= [ \text{Parent}(\text{pushed\_at}) - \text{Fork}(\text{pushed\_at}) ] \\
&< 90 \text{ days}
\end{split}
\label{eq:active_forks}
\end{equation}
\textit{\textbf{Identifying Stale forks:} } 
Potentially stale forks are branches that are not actively developed but may still hold some relevance to the main project. These forks might contain experimental features, bug fixes, or other contributions that could be valuable if integrated into the main codebase. Stale forks often contain code that could no longer be compatible with the current state of the original repository and do not contribute to the project's development. However, if these neglected forks are not regularly reviewed or maintained, the artifacts within them risk becoming unused and contributing to development waste. 

We apply the K-means clustering algorithm to classify these forks, setting the number of clusters to two instead of selecting other clustering algorithms. Our decision to exclude hierarchical, mean shift, gaussian mixture models (GMM), and self-organizing maps (SOM) from consideration is due to their inherent limitations of these clustering algorithms. 

The k-means statistical method partitions the forks into two clusters based on the difference in the \textit{`pushed\_at'} difference. The resulting clusters are labeled \textbf{`potentially stale'} and \textbf{`stale'}. The \textbf{`potentially stale'} are the forks that may not have recent activity but still hold potential for future updates or usage. In contrast, the \textbf{`stale'} cluster comprises forks that show a significant lack of activity, suggesting that they are unlikely to be updated. This classification not only aids in quantifying the extent of neglect among the forks but also helps understand the dynamics of repository maintenance and developer engagement in open-source projects. 

\subsection{Project Diversification Index (PDI) }
\label{subsection:project_diversification}
\textit{\textbf{Identifying Independent and contributing forks:}} Among the `active forks' from section \textit{Stale forks}, the \textit{project diversification index} helps identify the forks contributing to the project vs. the forks that are exploring directions currently out of the scope and vision of the parent project. 
\textit{Contributing forks} represent direct contributions to merge back into the parent project, reflecting active community participation. \textit{Independently developed forks} may diverge significantly from the original project, fulfilling distinct user requirements or exploring alternative directions. 

\begin{tcolorbox}
\textbf{Definition:} Measure of active work on copies of a project exploring their directions without contributing back to the parent project indicating a different vision.\\ 
\noindent
\textbf{SDW Type:} Building the Wrong Feature/Product \cite{sedano2017software}\cite{alahyari2019exploratory} 
\end{tcolorbox}

While Zhou et al. \cite{zhou2020has} used rules such as the description having \textbf{``fork of''} and having at least one year of development activity to identify forks into independently developing forks, these filters yielded a single fork out of all the forks of all the selected repositories in the section \ref{section:study_setup}. Additionally, these filters do not apply to more recently made repositories. Considering these observations and catering to a diverse category of repositories, the \textit{project diversification index} — a calculated measure aggregating pull requests and contributor engagement data. Using the PDI, we classify the forks into \textbf{`contributing forks'} and \textbf{`independently developed forks'} using a different set of rules. The PDI quantifies the extent to which active forks are either integrated with the parent project or are exploring their preferred direction that is not the parent project's direction, indicating a gap or misalignment between the customer need and the direction of the project, aka, \textbf{``building the wrong feature"}\cite{sedano2017software}. This index indicates a needed bridge between requirements addressed by the parent and the forked repositories, nudging towards \textit{building the wrong feature}. 
\subsection{PR Rejection Rate (PRR)}
\label{subsection:PR_rejection_rate}
Pull Requests (PRs) facilitate collaboration by allowing team members to provide feedback and approve modifications, ensuring that new additions meet the project's standards and requirements. An \textit{accepted PR} typically has undergone successfully passed code reviews, where other developers verify that the changes are error-free, align with the project's coding standards, and contribute positively to the functionality and objectives of the application. On the other hand, a \textit{PR rejection} can be because it introduces bugs, fails to comply with the established coding practices, or if the proposed changes do not align with the project's goals. Also, insufficient documentation or failure to pass automated testing can lead to PR rejection. 
The PR Rejection Rate indicates the proportion of work that does not make it into the project compared to the work that did, indicating `unused artifacts' as the amount of work that goes unaccepted and a need for `backlog management' to ensure maximum resource utilization. 
\label{subsection:PR_rejection_rate}
\begin{tcolorbox}
\textbf{Definition:} Measure the amount of work that does not get into the product.\\ 
\noindent
\textbf{SDW Type:} Mismanagement of Backlog \cite{sedano2017software}, Unused Artifacts  \cite{varanasi2024software} 
\end{tcolorbox}
\textit{\textbf{Calculating the PR Rejection Rate:}} We fetch all closed PRs and check if each had a `merged\_at' date, indicating the PR's changes were incorporated into the project. The ratio of unmerged to merged PRs gives the Unmerged-Merged Ratio. A lower ratio suggests effective project management, clear guidelines, and high-quality PR submissions that align with the project's needs. A higher ratio indicates inefficiency, with significant effort spent on unincorporated changes.
Higher Unmerged-Merged Ratios can result from contributors lacking understanding about project requirements due to poor communication with maintainers. This leads to submissions misaligned with project goals and not meeting necessary criteria, causing more PR rejections.

\subsection{Backlog Inversion Index (BI Index)}
\label{subsection:backlog_inversion}
Thorough backlog management helps prevent the accumulation of low-priority tasks that can clutter the backlog and delay essential updates. Backlog inversion occurs when a higher-priority issue or activity is added to the backlog before a lower-priority issue. Still, the lower-priority activity is closed before the higher-priority activity. This selection of a lower-priority task can happen for various reasons, from initially poor prioritization to subsequent re-prioritization of the backlog.
\begin{tcolorbox}
\textbf{Definition:} Working on lower priority items while higher priority items are in the backlog prior to the origin of the lower priority issue.\\ 
\noindent
\textbf{SDW Type:} Mismanagement of Backlog \cite{sedano2017software}, 
\end{tcolorbox}
\textit{\textbf{Measuring Backlog Inversion: }} By fetching all the issues, their creation, closure dates, and their priorities, we can identify the number of times the event of a lower priority task is completed before a higher priority task. The formula \ref{eq:bi} presents an event of a scenario when a backlog item of lower \textit{I2} priority is closed before a higher priority issue \textit(I1) that was in the backlog before the incoming lower priority issue. This total is normalized to the number of closed issues to calculate the BI Index. 
\begin{equation}
\label{eq:bi}
\frac{\sum_{\text{all events}} \left( (P(I_1) > P(I_2)) \land (C(I_2) = 1) \land (C(I_1) = 0) \right)}{\text{( 3 * total closed issues ) }}
\end{equation}

While backlog inversion in open-source software measures subjective choices and contributions, it indicates how contributors pick the issues. Hence, this measure can be used by teams in the industry to identify backlog inversion, potentially indicating the causes. 
\subsection{Feature Fulfilment Rate (FFR)}
\label{subsection:feature_fulfilment_rate}

Effective backlog management and closure are critical for maintaining the efficiency and success of software development projects. Metrics such as velocity, which measures the work a team can complete in a given iteration, are vital for tracking progress and making informed decisions about task prioritization. By monitoring velocity, teams can make informed decisions about task prioritization and resource allocation, preventing task accumulation. \cite{sedano2017software} identified the need for a balanced feature-to-bug ratio to reduce \textbf{`mismanagement of backlog'} \cite{sedano2017software}. Considering this proposal and project velocity, we propose the feature fulfillment rate. The FFR combines two types of graphs: the \textit{Feature-bug balance graph} and the \textit{inflow-outflow graph}. To present either of the graphs, we first fetch all the issues of the type \textbf{`bug' }and \textbf{`feature'} and their timestamps for the creation and closure events. 
\begin{tcolorbox}
\textbf{Definition:} Backlog Management practice visualizations to assist in understanding the current backlog health of the system in terms of `bugs' and `features' delivery.\\ 
\noindent
\textbf{SDW Type:} Mismanagement of Backlog \cite{sedano2017software}, 
\end{tcolorbox}
\textit{\textbf{Calculating the Feature Fulfilment Rate:} }  We classify the issues into bins of [0-5], [5-30], [30-90], and [90-180] days based on the number of days taken to close the issues. The [0-5] days bin focuses on capturing the ability of the team to quickly resolve urgent issues, which is crucial for maintaining high software quality and user satisfaction. We designed the [5-30] days bin to track somewhat urgent but not critical issues, indicating the team's effectiveness in handling routine and moderately complex tasks within a standard iteration cycle. The [30-90] days bin helps understand the team's capacity to manage and resolve more complex or less urgent issues requiring longer-term planning and execution. [90-180] days bin provides insights into the handling of long-standing issues, helping to identify and address if there are persistent problems in the backlog.

Under each bin, we present the normalized number of backlog items closed in the \textbf{`n'} number of days among the bins. \textit{Feature-Bug Balance Graph} shows the normalized number of issues closed during the day of the respective bins. This graph helps identify the balance between the prioritization of the \textit{`bugs'} and \textit{`features'} issue. A good balance between the \textit{`bugs'} and \textit{`features'} indicates a well-balanced and maintained backlog, hence lower `mismanagement of backlog.' Similarly, an imbalance between the closure of \textit{`bug'} and \textit{`feature'} points towards an ill-maintained backlog and suggests a need to re-prioritize issues. An acceptable trend would be a downward slope. 
\\
\textit{\textbf{Inflow-Outflow Graph}} presents the inflow and the outflow of the \textit{`bug'} and \textit{`feature'} issues, considering the possibility of spillovers from the previous sprints. The graph encompasses the past sixty sprints, where each sprint is fourteen days, which is the industry standard. This timeline covers just over two years of the project timeline. For each sprint, the graph shows the inflow and outflow. If the amount of items in the backlog during a sprint, including inflow and spillovers, is the same as the number of closed items, the graph would stand on 1. If the value is under 1, it indicates a higher inflow than the outflow, indicating \textit{`mismanagement of backlog'} and the need to re-prioritize it. On the positive side, when the value is higher than 1, additional efforts are put into the project to compensate for the existing items in the backlog. This combination of visualizations helps teams understand the overall project management of the bugs and features. The feature fulfillment rate quantifies the normalized closure duration required to resolve bugs and implement features within a software development process. This measure helps indicate and analyze the allocation efficiency between feature development and bug resolution in backlog management. An optimal balance between these two aspects suggests proficient backlog maintenance, thereby minimizing \textit{`mismanagement of backlog.'}

\section{Study Setup}
\label{section:study_setup}
GitHub API\footnote{https://docs.github.com/en/rest?apiVersion=2022-11-28} was used to fetch the details of the repositories on which the proposed measures are implemented \ref{section:measures}. An initial search on GitHub for repositories with over 100,000 stars yielded no results. Consequently, we adjusted the search criteria to target repositories with over 50,000 stars\footnote{https://github.com/search?q=stars\%3A\%3E50000\&type=Repositories} and select the top 50 owners with projects in English who have issues while excluding tutorials, resource collections, or consolidations. (for e.g., freecodecamp, ebookfoundation, interview prep). We excluded organizations that might skew the population, which in our case was Microsoft, which had over 5 thousand repositories compared to just over 3200 repositories from the rest of the organizations. We then applied the selection criteria on the repositories.\\ 
\noindent
\textit{\textbf{Inclusion Criteria:}} Pushed in 2024, has issues, has downloads stargazers > 0, forks > 0, started before 2024 for it should be mature. \\
\noindent
\textit{\textbf{Exclusion Criteria:}} Archived, is a template, fork. 

After sorting the repositories by the number of stars, we selected the top 10 repositories for our work. This list of chosen repositories is available in Table \ref{table:repository_selection}. 
\begin{table}[]
    \centering
    \caption{Repository Selection}
    \label{table:repository_selection}
    \begin{tabular}{c||c|c}
        \textbf{ID} &  \textbf{Organization} & \textbf{Repository} \\
        \hline
        \hline
        R0 & twbs & bootstrap \\
		R1 & facebook & react-native \\
		R2 & kubernetes & kubernetes \\
		R3 & axios & axios \\
		R4 & angular & angular \\
		R5 & rust-lang & rust \\
		R6 & puppeteer & puppeteer \\
		R7 & opencv & opencv \\
		R8 & rust-lang & rustlings \\
		R9 & vuejs & core \\	
        \hline
    \end{tabular}
    \vspace{1mm}
\end{table}

\section{Results}
\label{section:results}
\begin{figure*}[t]
\centering
\includegraphics[width=\textwidth]{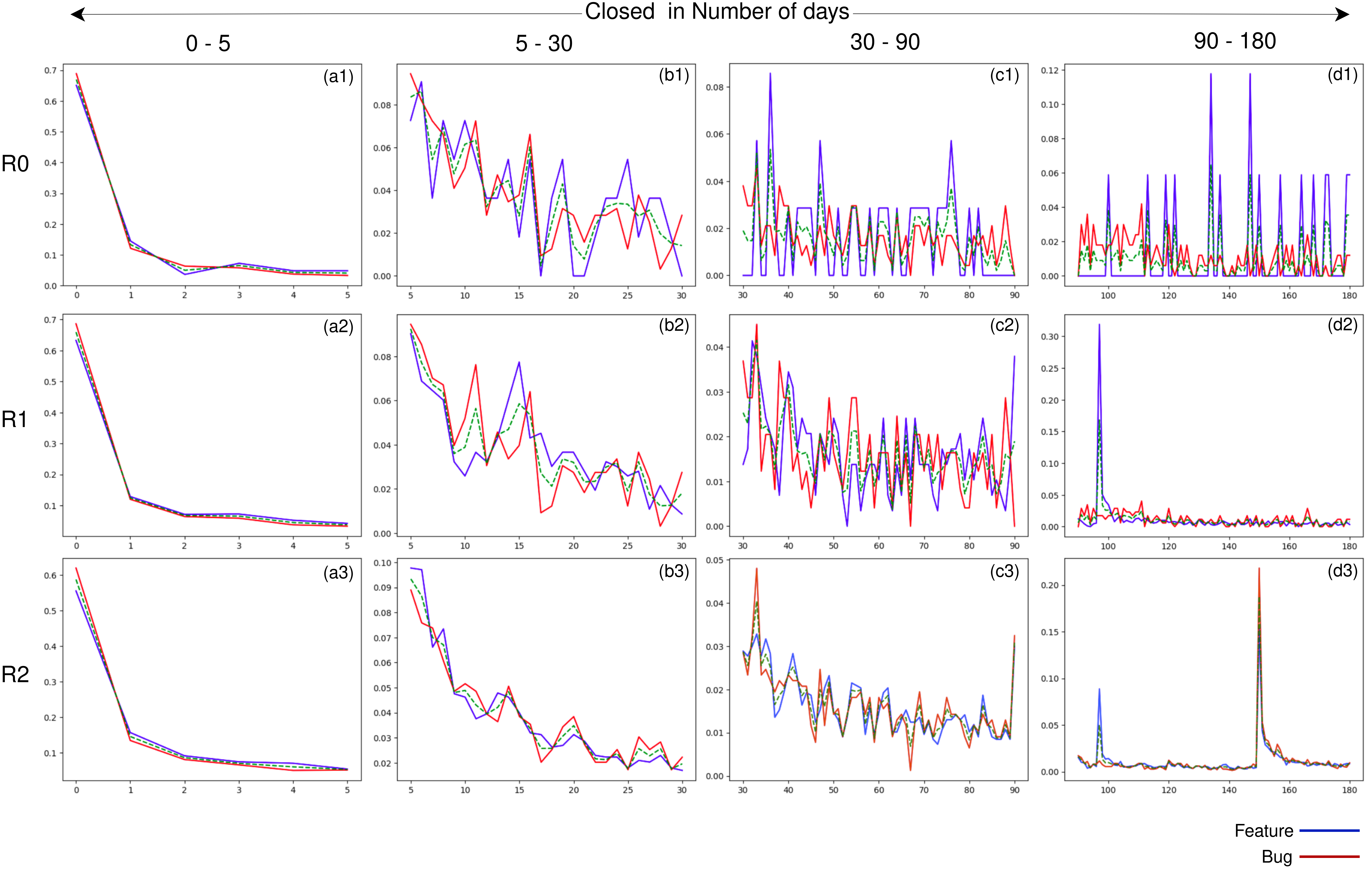}
\caption{Feature-bug balance\\ The x-axis is the ratio of the issues closed in the bin in the number of days since its inception shown in y-axis }
\label{fig:Feature_bug_balance}
\end{figure*}

\begin{figure*}[t]
\centering
\includegraphics[width=\textwidth]{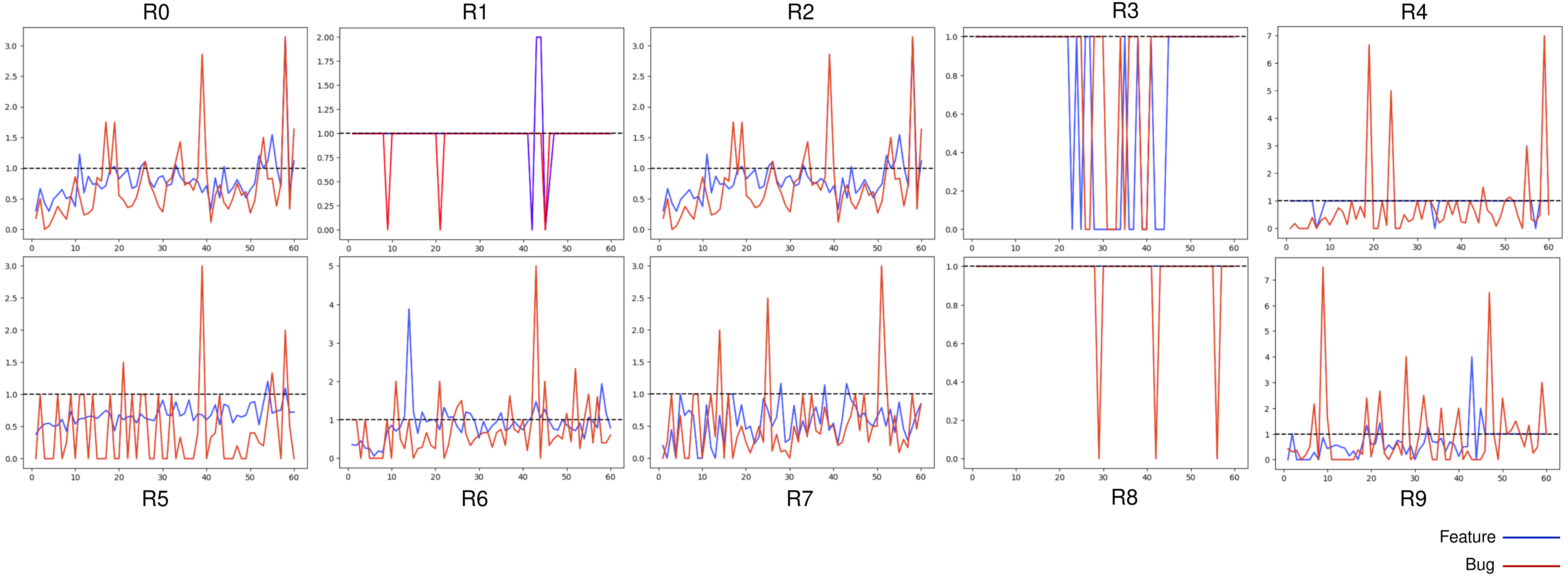}
\caption{Inflow-Outflow Graph 
\\ The x-axis represents the previous 60 sprints ending on April 30, 2024 with the y-axis portraying the backlog inflow to outflow management [ Value < 1 : number of issues closed during the sprint is less than the number of inflow and spillover issues; value > 1 : Compensating by closing issues that were behind; value = 1 : closing the same number of issues as inflow and spillover together for the sprint ]}
\label{fig:inflow_outflow_graph}
\end{figure*}

By applying the measures proposed in the section \ref{section:measures}, we present the results in this section. The code to implement the measures on a github repository is available in our resources \footnote{https://zenodo.org/records/11472533}. 
\\
\newline
\textbf{Defining Benchmarks:} 
We proposed measures as quantifiable value but not metrics. They captures waste traits via counts, percentages, ratios, calculations with an ideal target to reach closer to zero. Achieving zero is difficult in practice but can set a scale for organization using prior projects a threshold.
\\
\newline
\textbf{\textit{Stale Forks}}
The Table \ref{table:stale_forks_distribution} presents the counts and percentages of the classified forks into backup, active, potentially stale, and stale forks. A higher number on active forks is a good sign while a higher number on the stale and potentially stale forks is a bad sign. 
\begin{table}[]
    \caption{Fork Distribution}
    \label{table:stale_forks_distribution}
    \centering
    \begin{tabular}{c|c|c|c|c|c}
        \textbf{Project} & \textbf{Active} & \textbf{backup} & \textbf{Potentially} & \textbf{Stale} & \textbf{Total}\\
         &  &  & \textbf{Stale} & & \\
        \hline
        \hline
		R0 & 1,653 & 140,646 & 2,440 & 3,157 & 147,896 \\ 
		\% for R0 & 1.1176 & 95.0979 & 1.6498 & 2.1346 &  \\ 
		R1 & 1,331 & 36,550 & 1,707 & 2,576 & 42,164 \\ 
		\% for R1 & 3.1567 & 86.6853 & 4.0484 & 6.1094 &  \\ 
		R2 & 315 & 8,900 & 694 & 541 & 10,450 \\ 
		\% for R2 & 3.0143 & 85.1674 & 6.6411 & 5.1770 &  \\ 
		R3 & 275 & 18,692 & 738 & 456 & 20,161 \\ 
		\% for R3 & 1.3640 & 92.7136 & 3.6605 & 2.2617 &  \\ 
		R4 & 837 & 41,674 & 1,685 & 1,527 & 45,723 \\ 
		\% for R4 & 1.8305 & 91.1445 & 3.6852 & 3.3396 &  \\ 
		R5 & 951 & 12,738 & 2,937 & 1,766 & 18,392 \\ 
		\% for R5 & 5.1707 & 69.2583 & 15.9688 & 9.6020 &  \\ 
		R6 & 257 & 16,132 & 480 & 415 & 17,284 \\ 
		\% for R6 & 1.4869 & 93.3348 & 2.7771 & 2.4010 &  \\ 
		R7 & 902 & 101,316 & 2,027 & 1,522 & 105,767 \\ 
		\% for R7 & 0.8528 & 95.7916 & 1.9164 & 1.4390 &  \\ 
		R8 & 712 & 7,790 & 3,477 & 1,335 & 13,314 \\ 
		\% for R8 & 5.3477 & 58.5098 & 26.1153 & 10.0270 &  \\ 
		R9 & 318 & 11,938 & 1,030 & 564 & 13,850 \\ 
		\% for R9 & 2.2960 & 86.1949 & 7.4368 & 4.0722 &  \\ 
        \hline
    \end{tabular}
    \vspace{1mm}
\end{table}
\\
\newline
\textbf{\textit{Project Diversification Index: }}Among the active repositories, the Table \ref{table:project_diversification} shows the number of forks contributing back to the main project and those exploring their own directions. The ratio of these indicates the scope of directions that are not currently a part of the main project. The higher the PDI, the higher the waste \textit{`Building Wrong Feature/Product'}. 
\begin{table}[]
    \centering
    \caption{Project Diversification Index}
    \label{table:project_diversification}
    \begin{tabular}{c|c|c|c}
         &  & \textbf{Independently} & \textbf{Project} \\
        \textbf{Project} & \textbf{Contributing} & \textbf{developed} & \textbf{Diversification} \\
         & \textbf{forks} & \textbf{forks} & \textbf{Index} \\
        \hline
        \hline
		R0 & 23 & 1,605 & 0.0143 \\
		R1 & 132 & 1,066 & 0.1238 \\
		R2 & 64 & 187 & 0.3422 \\
		R3 & 14 & 246 & 0.0569 \\
		R4 & 114 & 609 & 0.1871 \\
		R5 & 387 & 177 & 2.1864 \\
		R6 & 22 & 213 & 0.1032 \\
		R7 & 74 & 753 & 0.0982 \\
		R8 & 32 & 648 & 0.0493 \\
		R9 & 58 & 202 & 0.2871 \\
        \hline
    \end{tabular}
    \vspace{1mm}
\end{table}
\\
\newline
\textbf{\textit{PR Rejection Rate: }}The Table \ref{table:PRrejectionRate} shows the number of PRs that have been merged and the number of PRs that were closed without a merge into the project. 
The higher the value, the higher the waste of \textit{`unused artifacts' }in the system as discussed in the section \ref{subsection:PR_rejection_rate}. 
\begin{table}[]
    \centering
    \caption{PR Rejection Rate}
    \label{table:PRrejectionRate}
    \begin{tabular}{c|c|c|c}
        \textbf{Project} & \textbf{Merged PRs} & \textbf{Unmerged PRs} & \textbf{Ratio} \\
        \hline
        \hline
		R0 & 8,712 & 6,302 & 0.7233 \\
		R1 & 816 & 15,795 & 19.3566 \\
		R2 & 58,194 & 20,381 & 0.3502 \\
		R3 & 825 & 632 & 0.7660 \\
		R4 & 3,549 & 23,409 & 6.5959 \\
		R5 & 54,913 & 16,570 & 0.3017 \\
		R6 & 4,488 & 1,108 & 0.2468 \\
		R7 & 12,050 & 2,732 & 0.2267 \\
		R8 & 861 & 505 & 0.5865 \\
		R9 & 2,447 & 1,550 & 0.6334 \\
        \hline
    \end{tabular}
    \vspace{1mm}
\end{table}
\\
\newline
\textbf{\textit{Feature fulfillment rate: }} \textbf{`Feature bug balance'} graphs of the repositories R0, R1, and R2 illustrate our results. Additional graphs are available as part our resources \footnote{https://zenodo.org/records/11472533}. The x-axis in figure \ref{fig:Feature_bug_balance} represents the number of days for closing an issue from its inception, and the y-axis represents the time it took for the issue to be closed in a number of days. The red line represents the normalized set of `bugs,' and the blue line is the \textbf{`features.'}. The graph \ref{fig:inflow_outflow_graph} shows the inflow of `bugs' and `features' over 60 sprints backtracked to end at April 30 2024, the month prior to final collection of the repository data for use in this study. In the figure \ref{fig:inflow_outflow_graph}, the red line represents `bugs', and the blue line represents the `features.' and the x-axis is the normalized ratio of the issues closed in the bin in the number of days since its inception shown in y-axis. 

\textbf{\textit{Backlog Inversion Index:}} The table \ref{table:backlog_inversion} consolidates the number of occurrences of the backlog inversion among the selected repositories. Ideally in a project, with good backlog management, we would want to have low to no backlog inversion. The \textit{High-Low, High-Medium and Medium-Low} are the possibility of the events and in turn the Backlog Inversion Index 
\begin{table}[]
    \centering
    \caption{Backlog Inversion}
    \label{table:backlog_inversion}
    \begin{tabular}{c|c|c|c|c}
         & \textbf{High} & \textbf{High} & \textbf{Medium} & \textbf{BI} \\
        \textbf{Project} & \textbf{Low} & \textbf{Medium} & \textbf{Low} & \textbf{Index} \\
        \hline
        \hline
		R0 & 0 & 0 & 0 & 0 \\
		R1 & 162 & 62 & 804 & 0.0137 \\
		R2 & 325153 & 68824 & 43699 & 3.3782 \\
		R3 & 0 & 6 & 0 & 0.0005 \\
		R4 & 551 & 115 & 21452 & 0.2837 \\
		R5 & 0 & 106043 & 0 & 0.8137 \\
		R6 & 10 & 9 & 126 & 0.0075 \\
		R7 & 1042 & 2708 & 40084 & 1.8132 \\
		R8 & 0 & 14 & 0 & 0.0089 \\
		R9 & 247 & 911 & 484 & 0.1306 \\
        \hline
    \end{tabular}
    \vspace{1mm}
\end{table}

\section{Discussion}
\label{section:discussion}
In this section we discuss our observations of the measures calculated on Open source software projects through our research questions from the Section 1.
\\
\newline
\textbf{RQ1: \textit{What are the known types of software development waste(s)? }}

Sedano et al. \cite{sedano2017software} and Alahyari et al. \cite{alahyari2019exploratory} conducted extensive participant-observation studies to compile a comprehensive list of SDWs. They commonly identified several types of waste, including building the wrong feature or product, rework, excessive cognitive load, multitasking, waiting time, and partially completed work. In addition, Sedano et al. \cite{sedano2017software} specifically identified wastes such as mismanagement of the backlog, unnecessarily complex solutions, psychological distress, knowledge loss, and ineffective communication. On the other hand, Alahyari et al. \cite{alahyari2019exploratory} highlighted wastes such as lengthy processes, management and organizational issues, and delayed releases.

Varanasi et al. \cite{varanasi2024software}, in their findings study on SDW during the COVID-19 pandemic, noted an amplification of wastes related to management and organizational aspects  previously identified by Alahyari et al. \cite{alahyari2019exploratory}. Varanasi et al. also reported that organizations experienced additional waste due to inadequate home office setups. Furthermore, both Alahyari et al. \cite{alahyari2019exploratory} and Varanasi et al. \cite{varanasi2024software} mentioned unused artifacts or pre-studies/proof of concepts that do not make it into the final product, as well as unnecessary processes and handovers. Apart from these studies, we have not found any other major contributions in this area. 
\\
\newline
\textbf{RQ2: \textit{How are these software development wastes measured?}}

Alahyari et al. observed that a significant portion of the industry still lacks formal measurement frameworks for waste. With 61\% of respondents in a study indicating no measures in place or a lack of waste measurement \cite{alahyari2019exploratory}. Instead productivity KPIs are used as proxy measures \cite{varanasi2024software}\cite{alahyari2019exploratory}. \textit{`Velocity'} and \textit{`lead time'} are metrics from agile project management that provide insights into team performance and process efficiency 
\cite{alahyari2019exploratory}\cite{petersen2012palette}. Additionally, cumulative flow diagrams are employed to visualize the flow of work across different phases of a software project and helps to identify process gaps \cite{petersen2012palette}. The table \ref{table:literature_measures} illustrate the observed measures and their definitions. 
\begin{table}[h!]
    \centering
    \caption{Measure(s) from Literature}
    \label{table:literature_measures}
    \begin{tabular}{m{0.25\columnwidth}|m{0.35\columnwidth}|m{0.15\columnwidth}}
        \textbf{Measure} & \textbf{Definition} & \textbf{Reference} \\
        \hline
        \hline
        Velocity & The rate at which a team or individual can deliver completed work items over a specific period of time. & \cite{alahyari2019exploratory} \\
        \hline
        Lead Time & The total time it takes for a task/project to be completed, from the moment it starts till it ends. & \cite{alahyari2019exploratory}\cite{petersen2012palette} \\
        \hline
        WIP (Work in Progress) & The number of tasks or projects that are currently in the process of being completed. & \cite{alahyari2019exploratory}\cite{petersen2012palette} \\
        \hline
        Defect Count & Defect inflow during a time period. & \cite{alahyari2019exploratory}\cite{petersen2012palette} \\
        \hline
        Trouble Reports & Descriptive reports of the. & \cite{alahyari2019exploratory} \\
        \hline
        Self-reporting & Self-reporting waste based on interview questions. & \cite{sallin2023waste} \\
        
    \end{tabular}
    \vspace{1mm}
\end{table}
\\
\newline
\begin{table}[]
    \centering
    \caption{Software Development Waste Measures}
    \label{table:SDWmeasures}
    \begin{tabular}{c|c|c|c|c}
         & \textbf{Stale \&} & \textbf{Project} & \textbf{Unmerged} & \textbf{Backlog} \\
        \textbf{} & \textbf{Potentially} & \textbf{Diversification} & \textbf{/ Merged} & \textbf{Inversion}  \\
        \textbf{} & \textbf{Stale \%} & \textbf{Index} & \textbf{Ratio} & \textbf{Index}\\
        \hline
        \hline
        R0 & 3.7844 & \textbf{0.0143} & 0.7233 & 0 \\
		R1 & 10.1579 & 0.1238 & \textbf{19.3566} & 0.0137 \\
		R2 & 11.8181 & 0.3422 & 0.3502 & \textbf{3.3782} \\
		R3 & 5.9223 & 0.0569 & 0.7660 & 0.0005 \\
		R4 & 7.0249 & 0.1871 & 6.5959 & 0.2837 \\
		R5 & 25.5709 & 2.1864 & 0.3017 & 0.8137 \\
		R6 & 5.1781 & 0.1032 & 0.2468 & 0.0075 \\
		R7 & 3.3554 & 0.0982 & 0.2267 & 1.8132 \\
		R8 & \textbf{36.1424} & 0.0493 & 0.5865 & 0.0089 \\
		R9 & 11.5090 & 0.2871 & 0.6334 & 0.1306 \\
        \hline
    \end{tabular}
    \vspace{1mm}
\end{table}
\vspace{-3pt}
\textbf{RQ3: \textit{How much software development waste(s) occurs in Open-Source software projects? }}
\\
\textbf{Feature \& Bug balance:} Table \ref{table:SDWmeasures} and Figures \ref{fig:Feature_bug_balance},  \ref{fig:inflow_outflow_graph} illustrates the overall observations that help us infer the occurrence of SDW in Open-source projects. Table \ref{table:SDWmeasures} presents that the repository R8 has higher stale forks than the rest. This high value represents the number of unused artifacts in the repository. The repository  R0 has the lowest project diversification index, suggesting that the diversification of forks is out of the current scope of the parent repository. Thus causing \textit{`building the wrong feature/product'} type of SDW in the repository R0. The repository R1 faces the most \textit{`unused artifacts'} SDW from the unmerged PRs, indicating the need for reinvestment into \textit{`backlog management'} for repository R0. The \textit{`backlog inversion index'} is highest in the repository R2, indicating the need for effective backlog management. 


Figure \ref{fig:Feature_bug_balance} shows graph for repositories R0, R1, and R2 with issue age over time. Graphs a1, a2, a3 are for R0, R1, R2 present issue age of 0-5 days. Most of these issues follow similar closure pattern. The overall 0-5 day bucket has steep drop indicating high resolution rate and shorter issue age. The graphs b1, b2, b3 for repositories R0, R1, R2 show 5-30 day issue age with more fluctuating trend for R0, R1 than R2. The trend stabilizes at lower level than initial 0-5 day bucket(s). This indicates repository R2 managed backlog more effectively than R0, R1.

The graphs c1, c2, c3 for repositories R0, R1, R2 present 30-90 day issue age bucket with higher peaks and valleys for R0, R1 than R2. R0 trend flats out. Thus shows prioritization of features after certain age with less consistent resolution timeframe. This varies significantly from day-to-day. The graphs d1, d2, d3 for repositories R0, R1, R2 present 90-180 day issue age bucket. We observe multiple peaks for R0, that shows backlog trigger at certain age. R0 feature peaks at 135, 150 days and R2 spike at 100 days for features which is similar to R0 at 100 days. The R0, R2 have high engagement at 150 days which shows that they revisit backlog frequently which aids steady project health.
Similarly, repository R2 shows a spike at the 100-day age for the features, which also happens to R0. Repositories R0 and R2 have additional mechanisms to create engagement with an issue closer to 150 days. Overall, revisiting the backlog frequently, with a plan suiting the project, will aid the project's steady health. 
\\
\newline
\textbf{Inflow-Outflow Data:} Figure \ref{fig:inflow_outflow_graph} presents the inflow and outflow of feature and bug issues along with respective spillovers for repositories R0, R1, R2, R3, R4, R5, R6, R7, R8 and R9. The inflow-outflow graphs of repositories R0, R2, R6, R7, and R9 show frequent fluctuations with higher spikes of bug issues. This observation illustrates that the contributors are compensating for their bug issues, leading to improper backlog management. Repositories R3 clearly show that the issues are steadily fixed within the subsequent sprints without causing an overload. In the case of R8, the bug issues are fixed instantaneously alongside feature issues within a given sprint.
In practice, we recommend open-source project owners to conduct a detailed waste analysis on closed projects while formulating the benchmarks for future projects especially for measures such as the feature fulfillment rate, PR rejection rate, and stale forks. They provide good insights into overall SDW across the open source project. 
\section{Limitations \& Threats to Validity}
\label{section:threats}
Calculating the proposed SDW measures requires access to source code repository metadata and historical version control data. The measures may yield insignificant insights if the underlying repository metadata is not continuously available. 
\noindent
\textbf{Internal validity:} A potential threat is the difference between the selected OSS repositories that could influence the results. To mitigate this, we used a repeatable and systematic method for repository selection, ensuring a representative sample of OSS projects. \textbf{External validity:} A possible threat is population validity, where the selected OSS repositories might not represent the broader population of OSS projects. As part of the selection process, we identified large projects that have been active for multiple years. Temporal validity, concerning the generalizability of findings to other time periods, was ensured by selecting repositories with consistent activity over significant periods. \textbf{Construct validity} involves the extent to which the measures accurately capture the concept of SDW in OSS projects. One threat is inadequate explanation of constructs, where ambiguity in defining "waste" could arise. We mitigated this by providing clear and detailed definitions of SDW, based on literature and expert input.  To minimize experimenter expectancies, where researchers' biases could influence measurements and interpretations, measures were applied in an automated and objective manner, and results were reviewed by multiple researchers. 
\section{Conclusion}
\label{section:conclusion}
SDW can negatively impact software product quality and delivery timelines. Early waste detection enables practitioners to avoid productivity issues and schedule delays. This paper proposes measures to identify SDW for adoption in real-time projects, validated using open-source projects as proof-of-concept. A strategy is outlined for industry practitioners to utilize these measures to improve productivity and meet delivery timelines more effectively. Future work aims to expand the measures to encompass other SDWs and collaborate with industry to evaluate the measures on domain-specific projects. This will facilitate understanding the practical impact and proposing new measures to address various types of SDW.

%% file: sample-sigconf.bbl

\begin{thebibliography}{12}


\ifx \showCODEN    \undefined \def \showCODEN     #1{\unskip}     \fi
\ifx \showDOI      \undefined \def \showDOI       #1{#1}\fi
\ifx \showISBNx    \undefined \def \showISBNx     #1{\unskip}     \fi
\ifx \showISBNxiii \undefined \def \showISBNxiii  #1{\unskip}     \fi
\ifx \showISSN     \undefined \def \showISSN      #1{\unskip}     \fi
\ifx \showLCCN     \undefined \def \showLCCN      #1{\unskip}     \fi
\ifx \shownote     \undefined \def \shownote      #1{#1}          \fi
\ifx \showarticletitle \undefined \def \showarticletitle #1{#1}   \fi
\ifx \showURL      \undefined \def \showURL       {\relax}        \fi
\providecommand\bibfield[2]{#2}
\providecommand\bibinfo[2]{#2}
\providecommand\natexlab[1]{#1}
\providecommand\showeprint[2][]{arXiv:#2}

\bibitem[Alahyari et~al\mbox{.}(2019)]%
        {alahyari2019exploratory}
\bibfield{author}{\bibinfo{person}{Hiva Alahyari}, \bibinfo{person}{Tony Gorschek}, {and} \bibinfo{person}{Richard~Berntsson Svensson}.} \bibinfo{year}{2019}\natexlab{}.
\newblock \showarticletitle{An exploratory study of waste in software development organizations using agile or lean approaches: A multiple case study at 14 organizations}.
\newblock \bibinfo{journal}{\emph{Information and Software Technology}}  \bibinfo{volume}{105} (\bibinfo{year}{2019}), \bibinfo{pages}{78--94}.
\newblock


\bibitem[Bufon and Leal(2019)]%
        {bufon2019method}
\bibfield{author}{\bibinfo{person}{M{\'a}rcio~Trov{\~a}o Bufon} {and} \bibinfo{person}{Adriano~Galindo Leal}.} \bibinfo{year}{2019}\natexlab{}.
\newblock \showarticletitle{Method for identification of waste in the process of software development in agile teams using lean and scrum}. In \bibinfo{booktitle}{\emph{Knowledge Management in Organizations: 14th International Conference, KMO 2019, Zamora, Spain, July 15--18, 2019, Proceedings 14}}. Springer, \bibinfo{pages}{466--476}.
\newblock


\bibitem[Gousios et~al\mbox{.}(2014)]%
        {gousios2014exploratory}
\bibfield{author}{\bibinfo{person}{Georgios Gousios}, \bibinfo{person}{Martin Pinzger}, {and} \bibinfo{person}{Arie~van Deursen}.} \bibinfo{year}{2014}\natexlab{}.
\newblock \showarticletitle{An exploratory study of the pull-based software development model}. In \bibinfo{booktitle}{\emph{Proceedings of the 36th international conference on software engineering}}. \bibinfo{pages}{345--355}.
\newblock


\bibitem[Hauge et~al\mbox{.}(2010)]%
        {hauge2010adoption}
\bibfield{author}{\bibinfo{person}{{\O}yvind Hauge}, \bibinfo{person}{Claudia Ayala}, {and} \bibinfo{person}{Reidar Conradi}.} \bibinfo{year}{2010}\natexlab{}.
\newblock \showarticletitle{Adoption of open source software in software-intensive organizations--A systematic literature review}.
\newblock \bibinfo{journal}{\emph{Information and Software Technology}} \bibinfo{volume}{52}, \bibinfo{number}{11} (\bibinfo{year}{2010}), \bibinfo{pages}{1133--1154}.
\newblock


\bibitem[Lin{\aa}ker et~al\mbox{.}(2023)]%
        {linaaker2023open}
\bibfield{author}{\bibinfo{person}{Johan Lin{\aa}ker}, \bibinfo{person}{Gregorio Robles}, \bibinfo{person}{Deborah Bryant}, {and} \bibinfo{person}{Sachiko Muto}.} \bibinfo{year}{2023}\natexlab{}.
\newblock \showarticletitle{Open Source Software in the Public Sector: 25 years and still in its infancy}.
\newblock \bibinfo{journal}{\emph{IEEE Software}} \bibinfo{volume}{40}, \bibinfo{number}{4} (\bibinfo{year}{2023}), \bibinfo{pages}{39--44}.
\newblock


\bibitem[Petersen(2012)]%
        {petersen2012palette}
\bibfield{author}{\bibinfo{person}{Kai Petersen}.} \bibinfo{year}{2012}\natexlab{}.
\newblock \showarticletitle{A palette of lean indicators to detect waste in software maintenance: A case study}. In \bibinfo{booktitle}{\emph{Agile Processes in Software Engineering and Extreme Programming: 13th International Conference, XP 2012, Malm{\"o}, Sweden, May 21-25, 2012. Proceedings 13}}. Springer, \bibinfo{pages}{108--122}.
\newblock


\bibitem[Rastogi and Nagappan(2016)]%
        {rastogi2016forking}
\bibfield{author}{\bibinfo{person}{Ayushi Rastogi} {and} \bibinfo{person}{Nachiappan Nagappan}.} \bibinfo{year}{2016}\natexlab{}.
\newblock \showarticletitle{Forking and the Sustainability of the Developer Community Participation--An Empirical Investigation on Outcomes and Reasons}. In \bibinfo{booktitle}{\emph{2016 IEEE 23rd international conference on software analysis, evolution, and Reengineering (SANER)}}, Vol.~\bibinfo{volume}{1}. IEEE, \bibinfo{pages}{102--111}.
\newblock


\bibitem[Sallin et~al\mbox{.}(2023)]%
        {sallin2023waste}
\bibfield{author}{\bibinfo{person}{Marc Sallin}, \bibinfo{person}{Martin Kropp}, \bibinfo{person}{Craig Anslow}, {and} \bibinfo{person}{Robert Biddle}.} \bibinfo{year}{2023}\natexlab{}.
\newblock \showarticletitle{Waste self-reporting for software development productivity improvement}. In \bibinfo{booktitle}{\emph{International Conference on Agile Software Development}}. Springer, \bibinfo{pages}{50--66}.
\newblock


\bibitem[Sedano et~al\mbox{.}(2017)]%
        {sedano2017software}
\bibfield{author}{\bibinfo{person}{Todd Sedano}, \bibinfo{person}{Paul Ralph}, {and} \bibinfo{person}{C{\'e}cile P{\'e}raire}.} \bibinfo{year}{2017}\natexlab{}.
\newblock \showarticletitle{Software development waste}. In \bibinfo{booktitle}{\emph{2017 IEEE/ACM 39th International Conference on Software Engineering (ICSE)}}. IEEE, \bibinfo{pages}{130--140}.
\newblock


\bibitem[Taibi(2015)]%
        {taibi2015empirical}
\bibfield{author}{\bibinfo{person}{Davide Taibi}.} \bibinfo{year}{2015}\natexlab{}.
\newblock \showarticletitle{An empirical investigation on the motivations for the adoption of open source software}. In \bibinfo{booktitle}{\emph{ICSEA 2015: the Tenth International Conference on Software Engineering Advances, November 15-20, 2015, Barcelona, Spain}}. IARIA, \bibinfo{pages}{426--431}.
\newblock


\bibitem[Varanasi et~al\mbox{.}(2024)]%
        {varanasi2024software}
\bibfield{author}{\bibinfo{person}{Dhiraj~SM Varanasi}, \bibinfo{person}{Sai~Anirudh Karre}, {and} \bibinfo{person}{Raghu Reddy}.} \bibinfo{year}{2024}\natexlab{}.
\newblock \showarticletitle{Software Development Waste amidst COVID-19 Pandemic: An Industry Study}. In \bibinfo{booktitle}{\emph{Proceedings of the 17th Innovations in Software Engineering Conference}}. \bibinfo{pages}{1--5}.
\newblock


\bibitem[Zhou et~al\mbox{.}(2020)]%
        {zhou2020has}
\bibfield{author}{\bibinfo{person}{Shurui Zhou}, \bibinfo{person}{Bogdan Vasilescu}, {and} \bibinfo{person}{Christian K{\"a}stner}.} \bibinfo{year}{2020}\natexlab{}.
\newblock \showarticletitle{How has forking changed in the last 20 years? a study of hard forks on github}. In \bibinfo{booktitle}{\emph{Proceedings of the ACM/IEEE 42nd International Conference on Software Engineering}}. \bibinfo{pages}{445--456}.
\newblock


\end{thebibliography}
